\documentclass[namedreferences]{solarphysics}

\usepackage[hyperref,optionalrh,showbiblabels]{spr-sola-addons} 
\usepackage{graphicx}        
\usepackage{color}           
\usepackage{breakurl}        

\chardef\us=`\_

\begin{document}

\begin{article}
\begin{opening}

\title{The New 2018 Version of the Meudon Spectroheliograph}

\author{J.-M.~\surname{Malherbe}$^{1}$\sep
        K.~\surname{Dalmasse}$^{2}$ }

\runningauthor{J.-M. Malherbe, K. Dalmasse}

\runningtitle{The new 2018 version of Meudon spectroheliograph}

\institute{$^{1}$ LESIA, Observatoire de Paris, 92195 Meudon, and
PSL Research University, France,
                 email: \url{Jean-Marie.Malherbe@obspm.fr}\\
              $^{2}$ IRAP, Universit\'{e} de Toulouse, CNRS, CNES, UPS, 31028 Toulouse,
              France,
               email: \url{Kevin.Dalmasse@irap.opm.eu}
               }

\begin{abstract}
Daily full-disk observations of the solar photosphere and
chromosphere started at Meudon Observatory in 1908. After a review
of the scientific context and the historical background, we describe
the instrumental characteristics and capabilities of the new version
operating since 2018. The major change is the systematic recording
of full line profiles over the entire solar disk providing 3D data
cubes. Spectral and spatial sampling are both improved. Classical 2D
images of the Sun at fixed wavelength are still delivered. We
summarize the different processing levels of on-line data and
briefly review the new scientific perspectives.
\end{abstract}
\end{opening}

\section{Introduction} \label{S-Introduction}

Systematic observations of the solar disk started at the Meudon
Observatory in 1908 with Deslandres spectroheliograph. The
collection is one of the largest available to the international
community with ten observed cycles (observations were only
interrupted during the First World War). It contains mainly
monochromatic images in wavelengths as H$\alpha$ (center),
Ca\textsc{ii} K1v (violet wing) and Ca\textsc{ii} K3 (center). A few
spectroheliographs were built at about the same time at Mount Wilson
\citep[USA:][]{Hal1,Ber}, Kodaikanal \citep[India:][]{Has} and
Coimbra \citep[Portugal:][]{Garcia}. The Meudon and Coimbra
spectroheliographs are still in use.

Most present full-disk observations of the Sun are performed with
narrow band imagers such as Fabry P\'{e}rot filters \citep[e.g.
Global Oscillation Network Group H$\alpha$ network:][]{Har} or Lyot
filters \citep[\emph{e.g.} Global H$\alpha$ Network:][]{Gal}.
Spectroheliographs, however, operate in spectroscopic mode and the
2018 release provides 3D data cubes ($x$, $y$, $\lambda$), $\lambda$
being the wavelength along line profiles and ($x$, $y$) the spatial
coordinates on the Sun.

This work is for the purpose of promoting the studies requiring
full-disk spectral data and/or long-term observations. Section 2
recalls the scientific interest of long-term archives in solar
physics. In Section 3, we present an historical background
summarizing the evolutions of the Meudon spectroheliograph from the
end of the 19th century up to now. Section 4 describes the
instrumental characteristics and capabilities of the 2018 version.
The data taken with the new spectrograph are widely open to the
community as explained in Section 5. Finally, Section 6 discusses
new perspectives.


\section{Scientific Context of Spectroheliograms and Long-Term Observations} \label{S-Scientific-Context}

The 110-year-old Meudon collection contains full-disk
spectroheliograms that are relevant to research on the solar cycle,
activity, and space weather. Ca\textsc{ii} K1v data make it possible
to analyze photospheric properties of active regions (ARs), while
H$\alpha$ reveals solar filaments. Such data are used to produce
synoptic maps of sunspots, faculae, and filaments \citep{Mar}, and
monthly publications \citep{Dazam} in \emph{l'Astronomie} during the
1928\,--\,2003 period. This in turn provides information on the
evolution of solar structures during the cycle, including the
equatorward migration of ARs \citep{Mak,Mou1,Cha} and the poleward
migration of high-latitude filaments \citep{Mou2,Kar}.

Sunspot number and area derived from full-disk Ca\textsc{ii} K1v
observations, as well as filament number and length derived from
H$\alpha$, further characterize the amplitude, duration, and
activity level of the solar cycle
\citep{Schwa,Wo,Hal2,Hoy,Hat1,Pet}. Quiescent H$\alpha$ filaments
delineate large cells of opposite magnetic polarities, which allowed
\cite{Leroy} to suggest different topologies of prominence magnetic
support. When analyzed over long periods, data series help to study
the cycle-to-cycle variability \citep[long-term modulations such as
the Gleissberg cycle or grand minima such as the Maunder minimum:
see the review by][]{Hat2}.

Long-term collections of full-disk observations provide constraints
to improve our understanding of the solar dynamo and its variations
\citep[\emph{e.g.}][]{Tob,Nor}. \cite{Mein3} used some structures of
spectroheliograms (spots and faculae) as magnetic tracers to
delineate large-scale motions. \cite{Ribes} found the presence of
zonal meridional circulation at several latitudes suggesting rolls
below the surface.

Full-disk Ca\textsc{ii} K1v and H$\alpha$ centennial archives are
also useful to study extreme solar flares such as the Carrington
event \citep{Ca,Hod} or the two-ribbon flare of 25 July 1946
\citep{El}. The largest sunspot region ever seen was registered on
05 April 1947, and it challenged researchers on the maximum size of
groups \citep{Aula1}. Studying extreme events requires long-term
archives that contribute to the development of prediction tools for
space weather applications \citep{Sch,Schm}.

Several works have proposed new proxies to extract information from
archives. \cite{Pev} developed a proxy to derive magnetic flux in
the plage and sunspots from Ca\textsc{ii} spectroheliograms. This
proxy will be important for pseudo-magnetogram reconstruction from
Ca\textsc{ii} data prior to 1974. It can be used to estimate the
amount of magnetic energy available for flares using scaling laws
derived from observations \citep{Cr,Sh,Kr,Par} or from MHD models
\citep{Aula1}. \cite{Toriu} defined a proxy to estimate the flare
energy from bright ribbons observed in H$\alpha$.

Long-term archives can suffer from various artifacts (\emph{e.g.}
over-exposure, aging and defaults, lack of photometric calibration,
missing temporal coverage). This can limit the analysis of the data
and the robustness of the results. \cite{Chat} recently proposed a
new method for automated processing and photometric calibration of
Ca\textsc{ii} spectroheliograms. This method makes it possible to
consistently combine and process images from different instruments
and observatories, in order to build datasets with better temporal
coverage. Such a method will be valuable for reconstructions of the
solar irradiance \citep{Fo,Lean} for investigating long-term
evolution and possible impact on Earth's climate \citep[see the
review by][]{Hai}.

In this context, the 2018 upgrade of the Meudon spectroheliograph
has two major goals:

\begin{description}
  \item[i)] improve the spectroscopic capabilities of the
  instrument (better spectral and spatial sampling, better wavelength
  centering of images)
  \item[ii)] develop the scientific potential by providing full
  line profiles over the entire Sun and adding new lines such as Ca\textsc{ii} H
  (later Ca\textsc{ii} infrared), allowing thermodynamical diagnostics
\end{description}


\section{Historical Background of the Meudon Spectroheliograph} \label{S-Historical-Context}

The rapid expansion of spectroscopy began 150 years ago. In 1868,
Janssen and Lockyer made the famous discovery of helium through
observation of the He\textsc{i} D3 587.5 nm line in the
chromosphere. In 1869, Young discovered the "coronium"
(Fe\textsc{xiv} green line identified 70 years later) in eclipse
spectra. These results \citep[see][for the detailed history]{Secchi}
initiated the development of solar spectroscopy, leading to the
spectroheliograph invention in 1892 by two astronomers in France and
USA \citep[][]{Hal1,Dela}.

The basic idea of spectroheliographs is to produce monochromatic
images of the solar photosphere and chromosphere using narrow
bandpass spectroscopic scans of strong Fraunhofer lines. For that
purpose, the Meudon spectroheliograph was characterized by:

\begin{itemize}
  \item a motorized objective (geographical E--W translation) moving the solar image over the
  entrance slit of the spectrograph
  \item dispersive elements (prisms and, later, gratings)
  \item a curved slit in the spectrum isolating the chosen spectral line
  \item a motorized photographic plate in the spectrum moving at the
  same speed as the imaging objective
  \item an occulting disk in the solar image to produce deep scans
  for prominences
\end{itemize}

The first full-disk spectroheliograms were obtained by Deslandres at
the Paris Observatory in 1894 using the Ca\textsc{ii} K line with
glass photographic plates (85.4 mm solar diameter). The first
long-exposure scans for prominences at the limb, using a disk
attenuator, were got at the same time. Deslandres moved from Paris
to Meudon in 1898 and a new laboratory and instrument were completed
in 1906, with the collaboration of \cite{Dazam}, who organized the
solar-observation routines.

\begin{figure}
\centering
\includegraphics[width=1.0\textwidth,clip=]{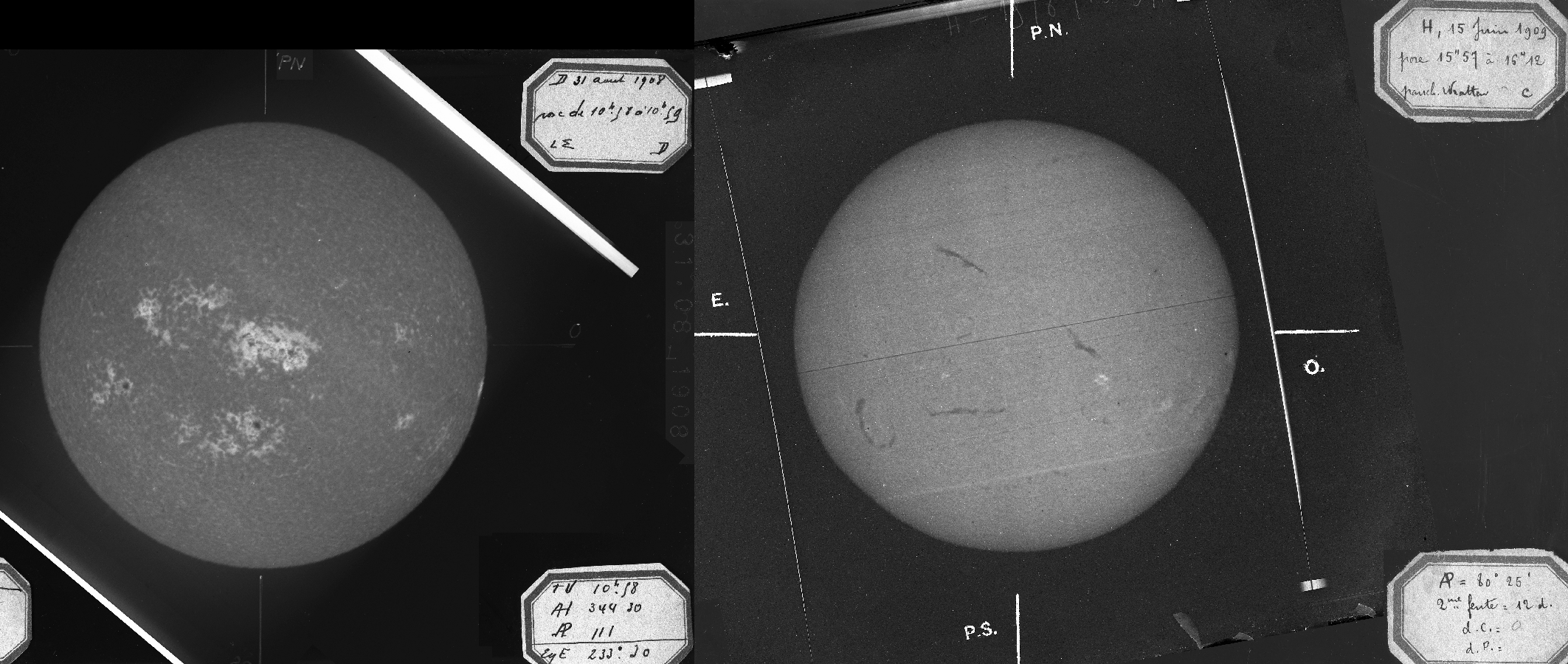}
 \caption[]{The Meudon collection of systematic Ca\textsc{II} K and H$\alpha$
 spectroheliograms started respectively in 1908 and 1909
 (here images of 31 August 1908, and 15 June 1909)
  } \label{kha}
\end{figure}

Systematic observations started in 1908 for Ca\textsc{ii} K and in
1909 for H$\alpha$ (Figure~\ref{kha}). The optics and mechanics were
modernized in 1985 by Olivieri. Photographic plates were abandoned
in 2001. The spectroheliograph no longer uses a selecting slit in
the spectrum, and the moving plate is replaced by a fixed electronic
detector. The 100,000 plates of the photographic collection are
progressively being digitized. Weather conditions in Meudon allow at
least one daily observation between 250 and 300 days per year.

\section{The 2018 Version of the Spectroheliograph} \label{S-New-Version}

\begin{figure}
\centering
\includegraphics[width=1.0\textwidth,clip=]{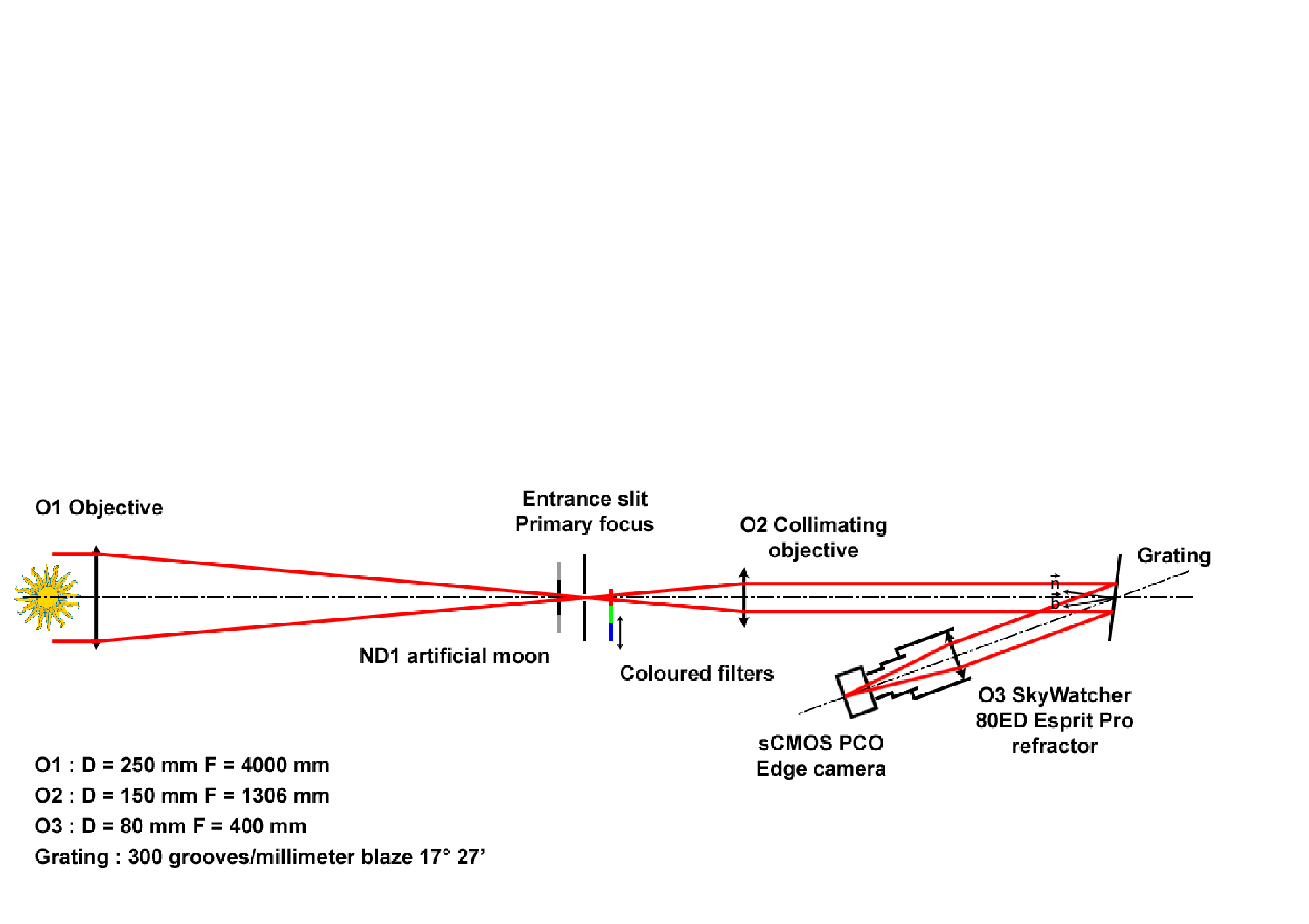}
 \caption[]{Optical design of the 2018 version of the spectroheliograh} \label{spectrohelio}
\end{figure}

As in previous versions, the solar light feeding the spectrograph
(Figure~\ref{spectrohelio}) is captured by a coelostat (two 400\,mm
flat mirrors) and directed horizontally towards a 250\,mm doublet
(O1, 4000\,mm focal length). It is optimized for the red (656\,nm)
and violet (394\,nm) part of the spectrum and forms a 37.2\,mm solar
image on the slit of the spectrograph. For other wavelengths, O1 can
be translated to adjust the focus. A diaphragm reduces the entrance
pupil to 170\,mm, providing a theoretical resolution of $1''$ in the
red part of the spectrum. In practice, this resolution is reduced to
$2''$ (or more) by local atmospheric turbulence.

The slit of the spectrograph has a width of 30 microns ($1.5''$ on
the Sun). The entrance objective O1 is motorized, so that the solar
image moves slowly in the geographic E--W direction (scanning the
full Sun typically takes 60 seconds). This motion decreases the
resolution from $1.5''$ to $2''$, but there is no consequence on
image quality, due to seeing conditions.

The focal length of the collimator O2 is 1300\,mm. This doublet is
fixed and optimized for 656 and 394\,nm; for other wavelengths, we
insert into the beam a compensating plate.

The dispersion device is a grating (300 grooves/mm and
$17^{\circ}27'$ blaze angle). Incident and diffraction angles are
respectively $7^{\circ}$ and $27^{\circ}$. The interference order
(3, 4, 5) is selected by filters behind the image plane
(respectively 650, 500, 400 nm CWL and 78, 55, 66 nm FWHM). The
wavelength switching is automatic and takes one minute.

In the present instrument, the camera lens O3 is an achromatic
SkyWatcher 80ED, 400\,mm focal length, forming a 11.4\,mm image on
the detector. The new camera uses a Fairchild 2020 scientific CMOS
sensor with rolling electronic shutter. It is water cooled at
$5^{\circ}$ C. This is a low-noise device with 30,000 dynamic range
and 16 bits digitization (0.5 electron per count). The format is
2048 $\times$ 2048 square pixels (6.5 microns, 30,000 electrons
full-well capacity). High speed readout (100 frames per second)
allows fast spectroscopy. The quantum efficiency is for
Ca\textsc{ii} H/K, H$\beta$, and H$\alpha$ respectively 20\,\%,
65\,\%, and 70\,\%.

The spectral range is line-dependant. It usually varies from 40 to
100 pixels (typically 0.5 nm to 1.0 nm, see Table 1) and can be much
enlarged (up to 10 nm) for specific programs upon request. The Sun
is scanned using 2048 steps. The final angular pixel is $1.1''$
(half of the best seeing at Meudon) with 3\,\% seasonal fluctuation
in km.

The exposure time is chosen to work at half saturation in the far
line wings (photon S/N ratio 100). It varies from 10 to 100\,ms
depending on line, disk attenuator, weather conditions, and height
of the Sun. 10\,ms is typically used for H$\alpha$, 20\,ms for
Ca\textsc{ii} H/K and 50\,ms for prominences with disk attenuator
(20 to 100 seconds scan duration). The speed of objective O1 is
automatically adjusted to the exposure time.

Two kinds of observations are performed daily:
\begin{description}
  \item[i)] full-disk observations with short exposure time
  \item[ii)] prominence observations with solar disk attenuated by an
  occulter of neutral density 0.9 (13\,\% transmission) in the image plane
  (this avoids disk saturation of long exposure scans)
\end{description}

\begin{figure}
\centering
\includegraphics[width=1.0\textwidth,clip=]{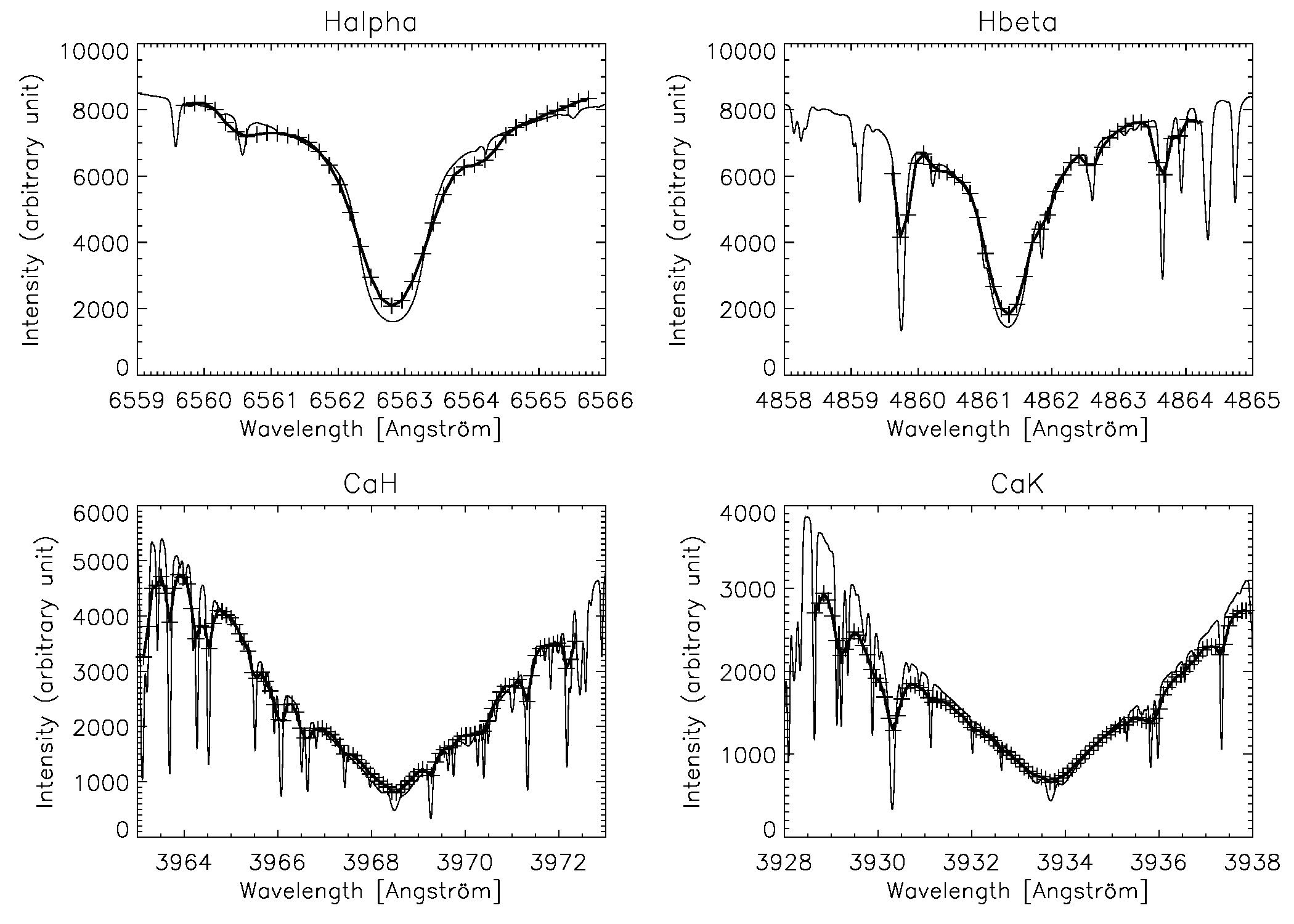}
 \caption[]{H$\alpha$, H$\beta$, Ca\textsc{ii} H and Ca\textsc{ii} K line profiles near disk center (\emph{thick line}) in comparison
 with atlas profiles, from \cite{Del} (\emph{thin line}). Sampling points are indicated by crosses.} \label{4profils}
\end{figure}

H$\alpha$, H$\beta$, and Ca\textsc{ii} are observed successively
(but Ca\textsc{ii} H and K are simultaneous) in interference orders
3, 4, and 5. Figure~\ref{4profils} shows typical spectra
$P_{\textnormal{obs}}(\lambda)$ obtained near disk center. The
comparison with the atlas $P_{\textnormal{ref}}(\lambda)$ by
\cite{Del} recorded at the Jungfraujoch observatory (3600\,m) with
exceptional spectral resolution (0.2\,pm) shows that lines are
smoothed, as the spectroheliograph is designed for broad lines. A
linear fit in the form $P_{\textnormal{obs}}(\lambda)= a + b \times
P_{\textnormal{ref}}(\lambda)$ reveals in line cores a small amount
of stray or parasitic light (a few \%).

\begin{table}
\begin{tabular}{cccccc}
  \hline
  Line & Wavelength & Order & Spectral Pix. & Spectral Res. & Spectral field \\
       & [nm]       &       & [nm]           & [nm]          & [pixels]       \\
  \hline
  H$\alpha$ & 656.28 & 3 & 0.0155 & 0.025 & 40  \\
  H$\beta$  & 486.13 & 4 & 0.0116 & 0.019 & 40  \\
  Ca\textsc{ii} H    & 396.85 & 5 & 0.0093 & 0.015 & 100  \\
  Ca\textsc{ii} K    & 393.37 & 5 & 0.0093 & 0.015 & 100  \\
  \hline
\end{tabular}
\caption{List of observable lines}
\end{table}

MP4 movies are included as Electronic Supplemental Material to
illustrate the principle of the 2018 version of the
spectroheliograph (see the Appendix for details). At the moment,
H$\beta$ is not yet operating, so that daily observations consist of
one or several series of four sequences including:

\begin{description}
  \item[i)] H$\alpha$ full-disk (short exposure)
  \item[ii)] H$\alpha$ prominences + disk attenuator (long exposure)
  \item[iii)] Ca\textsc{ii} H and K full-disk (short exposure)
  \item[iv)] Ca\textsc{ii} H and K prominences + disk attenuator (long exposure)
\end{description}

\begin{figure}
\centering
\includegraphics[width=1.0\textwidth,clip=]{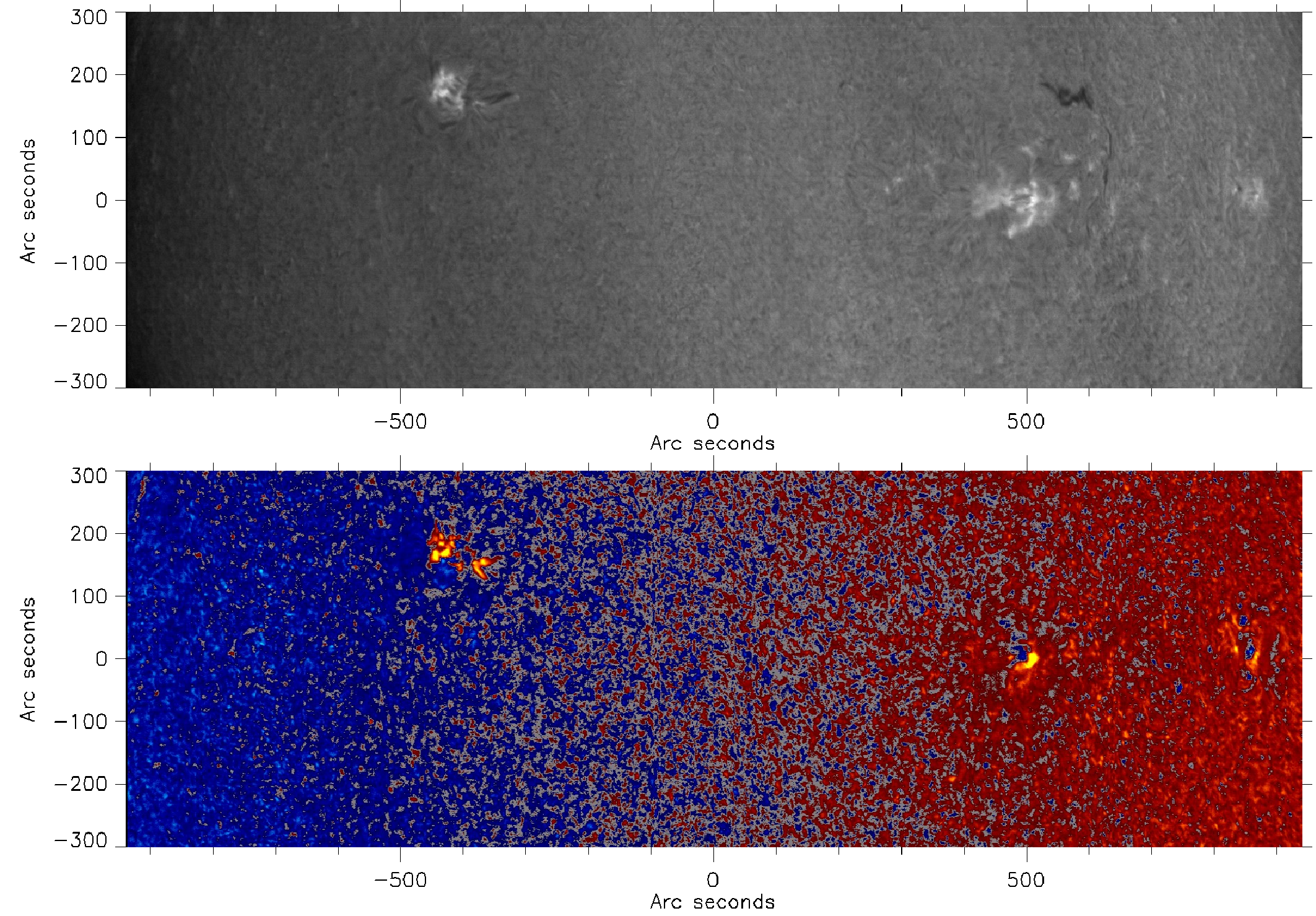}
 \caption[]{H$\alpha$ intensitygram (top) and dopplergram (bottom) showing solar rotation, AR12715 (left) and AR12713 (right), 20 June 2018, 08:44 UT.
 Intensities are not uniform because of transparency variation.} \label{dop}
\end{figure}

H$\alpha$ line profiles allow the computation of Dopplergrams, as
shown by Figure~\ref{dop}. We used a gaussian fit to detect the
wavelength position of the line, but other methods working well with
similar spectral resolutions may be used, as the bisector applied by
\cite{Mein1} to multichannel subtractive double pass observations.

\section{Data Levels and Dissemination} \label{S-Data-Levels}

We offer to the solar community two data levels:

\begin{description}
  \item[i)] Level 0: raw data cubes, in TIFF format available via anonymous FTP
  and dispatched in monthly catalogues at:

  \url{ftp://ftpbass2000.obspm.fr/pub/meudon/spc/tiff/}

  \item[ii)] Level 1: data issued from a standard procedure and delivered
  by the solar database at:

  \url{http://bass2000.obspm.fr/home.php?lang=en}
\end{description}


\subsection{Level 0 Data} \label{S-Level0}

Level 0 data are 3D raw data containing 2048 planes of spectral
images ($x$, $\lambda$) resulting from the solar scan in the
y-direction, except for the dark currents (DC), which are 2D images.
DC can be directly subtracted from each plane of the 3D raw
observations.

More technical details and the IDL code to read raw data and reorder
coordinates to produce ($x$, $y$, $\lambda$) datacubes are available
here:

\url{http://observations-solaires.obspm.fr/Meudon-spectroheliograph-2018-version}


\subsection{Level 1 Data} \label{S-Level1}

Level 1 is the result of a standard processing applied to level 0
data, including DC subtraction, correction of the spectral line
curvature (parabolic fit), rotation (P-angle and coelostat position)
in order to present the solar North at the top of the image.

Level 1 delivers FITS 3D output containing full line profiles ($x$,
$y$, $\lambda$), classical full-disk images (FITS for scientific
purpose and quick look JPEG) and a set with superimposition of
heliographic grids or polar angle graduations.

The optical procedure to produce flat fields is complicated and
under development. It will be applied in Level 1 when available.
Turbulence effects vary along the scan. Also, as the scanning and
spectra acquisition speeds cannot be exactly equal, a residual
distorsion between geographic N--S and E--W may exist. It could be
corrected using the method proposed by \cite{Mein3}.

\begin{figure}
\centering
\includegraphics[width=1.0\textwidth,clip=]{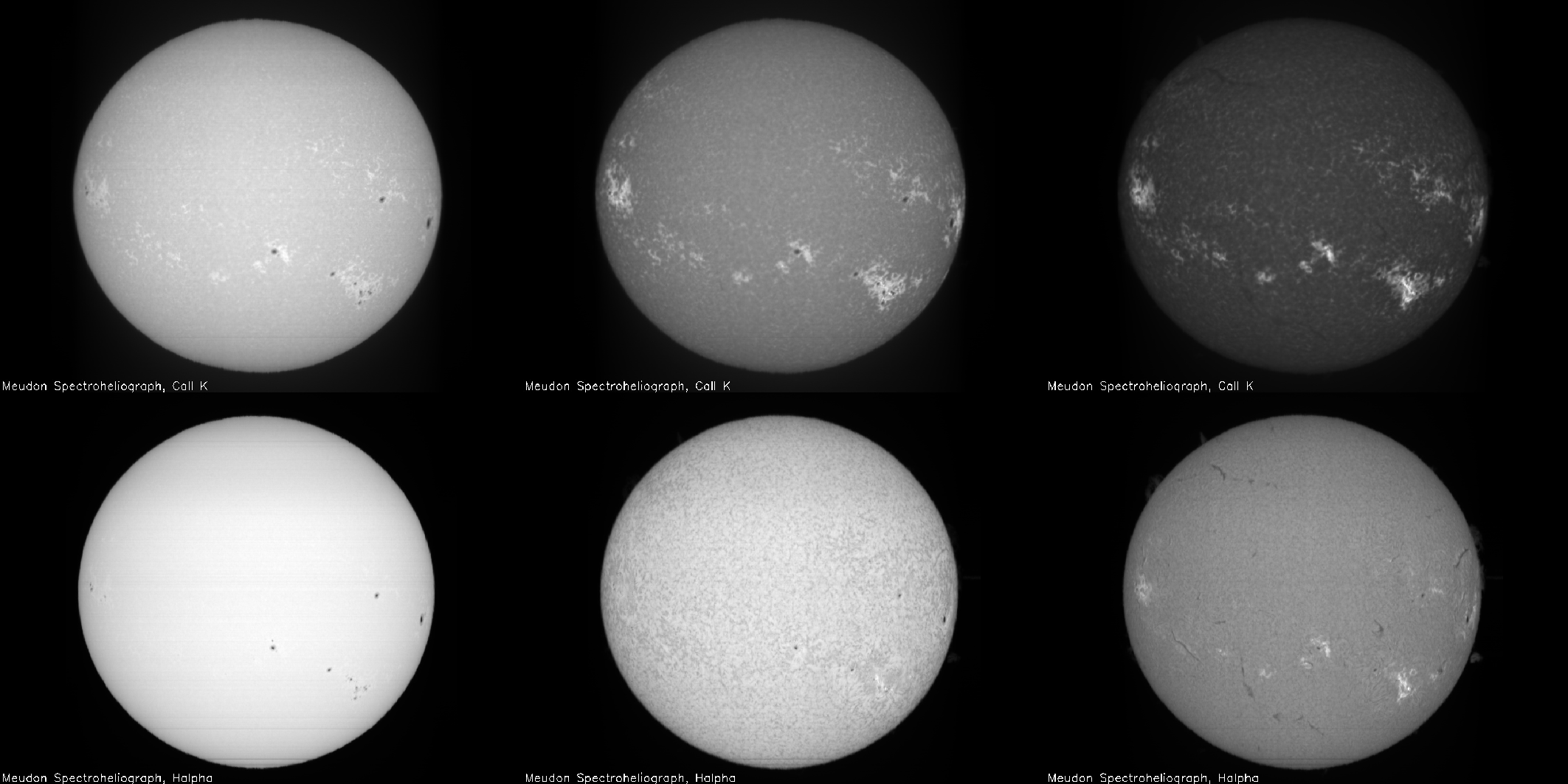}
 \caption[]{Example of images derived from profiles. \textbf{Bottom}: H$\alpha$ -0.15\,nm, -0.05\,nm,
 center. \textbf{Top}:  Ca\textsc{ii} K -0.15\,nm, -0.05\,nm,
 center (tests of November 13, 2013)} \label{images}
\end{figure}

Classical images (Figure~\ref{images}) usually derived from line
profiles are:
\begin{description}
  \item[i)] Ca\textsc{ii} H3 and K3 line core (full-disk and prominences with attenuator)
  \item[ii)] Ca\textsc{ii} H1v and K1v violet wing (-0.15\,nm apart from line center)
  \item[iii)] H$\alpha$ line core (full-disk and prominences with attenuator)
  \item[iv)] H$\alpha$ line wings (-0.05 and +0.05\,nm apart from line center)
  \item[v)] H$\alpha$ blue continuum (-0.15\,nm apart from line center)
\end{description}

\section{Discussion} \label{S-Discussion}

The Meudon spectroheliograph is designed for full-disk observations
of broad lines such as hydrogen or calcium. The characteristics of
successive versions are summarized in Table 2.

\begin{table}
\begin{tabular}{lcccccc}
  \hline
  Version & Spatial  & Spectral & Spectral pix.  & Dynamic & O3 & output \\
          & pixel    & FOV      & CaK-H$\alpha$ & range   & f   & slit       \\
          & [arcsec] & [pixels] & [nm]             & [half sat.] & [mm] & [$\mu$] \\
  \hline
  PHOTO 1908\,--\,2001   & 1.5      & 1         & 0.015\,--\,0.025 & density 3 & 3000 & 75 \\
  CCD 2002\,--\,2017     & 1.6      & 5         & 0.015\,--\,0.025 & 7500 & 900 & \\
  SCMOS 2018\,--\,       & 1.1      & 40\,--\,100 typical & 0.009\,--\,0.015 & 15000 & 400 & \\
                      &          & 1000 max &   &   & &
                      \\
  \hline
\end{tabular}
\caption{Spectroheliograph versions}
\end{table}

The photographic version (image diameter 85.4\,mm, 1908\,--\,1981 on
10 $\times$ 13\,cm glass plates and 1982\,--\,2001 on 13 $\times$
18\,cm film plates) is affected by defaults that are discussed by
\cite{Mein3}. The curved slit in the spectrum (75 microns width) was
removed with the first CCD (2002\,--\,2017) providing 5 wavelengths
around line core. With this shutterless frame transfer CCD, a small
amount of parasitic light differentially affects the five
wavelengths during the transfer. Full line profiles are available
since 2018 with a fast sCMOS sensor, which suppresses this
phenomenon and has better spectral resolution and improved dynamic
range. Scanning distortions still exist.

The 2018 upgrade allows the production of monochromatic images that
are more precisely centered in wavelength, with smaller bandwidth,
and better spatial sampling than before. Wavelength centering is now
done by computer, while in the former version, it was by visual
inspection. Line inclination and curvature are now corrected (Level
1 data only).

The availability of full line profiles is the major change since
1908. It is now possible to derive dopplergrams of active regions in
H$\alpha$. The spectral range is broad (0.5\,nm or more), so that
large velocities can be recorded. Filament tracking, which may be
affected by Doppler shifts, is improved. The spectroheliograph can
follow dynamic events such as flares or eruptive filaments, with
moderate temporal resolution (a few minutes) in the case of
full-disk scan. A faster mode is implemented for active regions
(reduced FOV). However, continuous observations cannot be
systematic, because of personnel limitations and meteorological
conditions, but participation in flare campaigns is possible.

Many parameters (temperature, velocity, and their gradients) can be
extracted from Ca\textsc{ii} line profiles. \cite{Mein2} proposed a
non-linear method based on Fourier series. Disturbances of Fourier
coefficients were compared to theoretical disturbances of
atmospheric models. Corresponding line profiles were computed using
a NLTE radiative transfer code. Temperature and velocity
fluctuations, together with vertical gradients, were derived from a
least-square inversion method. It was applied successfully to a time
sequence of Ca\textsc{ii} 396.8 nm line. Blue peaks occurring in
profiles were interpreted as downward velocity gradients associated
with temperature enhancement. \cite{Mol} used another NLTE code to
build a grid of H$\alpha$ lines and inverted observed profiles of
filaments. \cite{Mein4} made a cloud model to fit Ca\textsc{ii}
854.2 nm profiles also in the case of filaments. More recently,
\cite{Bec} presented a fast inversion code for Ca\textsc{ii} 854.2
nm. It it based on an archive of theoretical spectra that are
computed under the assumption of LTE. From a comparison to NLTE
inversions, they applied a first-order correction to the parameters
deduced from the LTE case. \cite{Ku} also produced inversions of the
Ca\textsc{ii} 854.2 nm line using a NLTE code to investigate the
evolution of temperature and velocity in the flaring chromosphere.

For these reasons, systematic observation of Ca\textsc{ii} infrared
lines are planned for the Meudon spectroheliograph.


\section{Conclusion} \label{S-Conclusion}

The 2018 upgrade of the Meudon spectroheliograph marks a major
change since 1908. It provides full line profiles over the entire
solar disk, with good spectral sampling (0.010\,nm to 0.015\,nm)
over a 1\,nm typical spectral field (up to 10\,nm upon request). 3D
data cubes ($x$, $y$, $\lambda$) are now produced daily, together
with classical images at different wavelenths along the profiles.
H$\alpha$, Ca\textsc{ii} H, and K lines are routinely observed;
H$\beta$ and Ca\textsc{ii} infrared lines will come later. Both
Level 0 (raw) and 1 data are available on-line to the community.

Spectra observed with the spectroheliograph are good indicators of
solar activity in the chromosphere (active regions, faculae,
filaments, prominences), while activity in the transition region and
corona is provided by Solar Dynamics Observatory (Atmospheric
Imaging Assembly, AIA). The 160\,and 170\,nm continua observed by
AIA at the top of the photosphere give images close to Ca\textsc{ii}
(except for filaments), so that combined ground-based and space
observations produce a complete diagnostic of the solar atmosphere.

The main scientific objective of the Meudon spectroheliograph, with
a few series of daily observations, concerns long-term solar
activity, rare and extreme events, and the continuation of the ten
solar cycles collection. Ca\textsc{ii} retraces the history of the
surface covered by magnetic fields, and H$\alpha$ records
chromospheric phenomena. Line profiles open new perspectives in
terms of temperature and velocity determination by inversion
techniques.

Data before 2002 (images in Ca\textsc{ii} K3, K1v, and H$\alpha$)
are registered on photographic media (glass or film) and scanned. We
intend to recover the production of synoptic maps using automated
procedures \citep{Abou}.

A complementary short-term and automatic survey at high temporal
resolution (20 seconds) for observations of transient activity
(fares, filament eruption, CME onset, Moreton waves) is under
development for the Calern Observatory using monochromatic H$\alpha$
and Ca\textsc{ii} K filters; this will be the topic of a later
article.

\vspace{0.5cm}

\emph{If you use data for publication, please insert the following
acknowledgment}: ``Meudon spectroheliograph data are courtesy of the
solar and BASS2000 teams, as part of operational services of Paris
Observatory''

\begin{acks}

This article is dedicated to the memory of E. Ribes who directed
Meudon solar observations during many years and passed away
prematurely in 1996.

\vspace{0.25cm}

We thank the referee for helpful comments and suggestions. We are
also indebted to the technical staff of LESIA (D. Crussaire, N.
F\"{u}ller, C. Imad, M. Ortiz, C. Reni\'{e} and D. Ziegler) and to
the team of observers which operate daily the instrument (C.
Blanchard, I. Bual\'{e}, S. Cnudde, A. Docclo, I. Ibntaieb). We are
grateful to the Scientific Council of Paris Observatory for
financial support and INSU/CNRS for the label of National Observing
Service (SNO). K.D. is supported by the Centre National d'Etudes
Spatiales (CNES).
\end{acks}

\section*{Disclosure of Potential Conflicts of Interest}

The authors declare that they have no conflicts of interest.

%
\appendix

Electronic Supplemental Material (movies in MPEG 4 format).

\begin{description}

\item[i)] Movie 1: the principle of the 2018 upgrade of the Meudon spectroheliograph in
the case of Ca\textsc{ii} K line. The surface of the Sun (right) is
scanned by a thin slit and full line profiles (left) are recorded
for the 2048 positions of the slit. Data are stored in 3D files (2
space coordinates + spectral line). The scan time takes about 60
seconds using a $1.5''$ slit.

\item[ii)] Movie 2: same as Movie 1, but for H$\alpha$ line.

\item[iii)] Movie 3: example of images got at different wavelength positions
along the line profiles. Part 1: H$\alpha$ ; Part 2: Ca\textsc{ii}
K.

\end{description}

\bibliographystyle{spr-mp-sola}
\bibliography{JMM_sola_V4}

\IfFileExists{\jobname.bbl}{} {\typeout{}
\typeout{****************************************************}
\typeout{****************************************************}
\typeout{** Please run "bibtex \jobname" to obtain} \typeout{**
the bibliography and then re-run LaTeX} \typeout{** twice to fix
the references !}
\typeout{****************************************************}
\typeout{****************************************************}
\typeout{}}

\end{article}

\end{document}